\documentclass[aps,prd,amsmath,floats,floatfix, twocolumn,
superscriptaddress,nofootinbib,showpacs,longbibliography]{revtex4-1}
\usepackage[T1]{fontenc}
\usepackage[utf8]{inputenc}
\usepackage[english]{babel}
\usepackage{dcolumn}
\usepackage{bm}
\usepackage{graphicx}
\usepackage[dvipsnames, usenames]{xcolor}
\usepackage{float}
\usepackage[normalem]{ulem}
\usepackage{bbold}
\definecolor{linkcolor}{rgb}{0.0,0.3,0.5}
\usepackage[hypertexnames=false, unicode, colorlinks=true, linkcolor=linkcolor,
citecolor=linkcolor, filecolor=linkcolor,urlcolor=linkcolor,
pdfusetitle]{hyperref}

\newcommand{\av}[1]{{\textcolor{purple} {{}}}}

\usepackage{listings}
\usepackage{orcidlink}

\usepackage{xparse}
\usepackage{booktabs}

\begin{document}

\selectlanguage{english}

\keywords{<keywords>}

\title{Rapid Identification and Classification of Eccentric Gravitational Wave Inspirals with Machine Learning}

\author{Adhrit Ravichandran \orcidlink{0000-0002-9589-3168}}
\affiliation{International Centre for Theoretical Sciences, Tata Institute of Fundamental Research, Bangalore 560089, India}
\affiliation{Center for Scientific Computing and Data Science Research, University of Massachusetts, Dartmouth, MA 02747, USA}
\author{Aditya Vijaykumar \orcidlink{0000-0002-4103-0666}}
\affiliation{International Centre for Theoretical Sciences, Tata Institute of Fundamental Research, Bangalore 560089, India}
\affiliation{Department of Physics, University of Chicago, Chicago, IL 60637, USA}
\author{Shasvath J. Kapadia \orcidlink{0000-0001-5318-1253}}
\affiliation{The Inter-University Centre for Astronomy and Astrophysics, Post Bag 4, Ganeshkhind, Pune 411007, India}
\affiliation{International Centre for Theoretical Sciences, Tata Institute of Fundamental Research, Bangalore 560089, India}
\author{Prayush Kumar \orcidlink{0000-0001-5523-4603}}
\affiliation{International Centre for Theoretical Sciences, Tata Institute of Fundamental Research, Bangalore 560089, India}

\begin{abstract}
 Current templated searches for gravitational waves (GWs) emanated from compact binary coalescences (CBCs) assume that the binaries have circularized by the time they enter the sensitivity band of the LIGO-Virgo-KAGRA (LVK) network. However, certain formation channels predict that in future observing runs (O4 and beyond), a fraction of detectable binaries could enter the sensitivity band with a measurable eccentricity $e$. Constraining $e$ for each GW event with Bayesian parameter estimation methods is computationally expensive and time-consuming. This motivates the need for a machine learning based identification and classification scheme, which could weed out the majority of GW events as non-eccentric and drastically reduce the set of candidate eccentric GWs. As a proof of principle, we train a separable-convolutional neural network (SCNN) with spectrograms of synthetic GWs added to Gaussian noise characterized by O4 representative \texttt{PSD}s. We use the trained network to (i) segregate candidates as either eccentric or non-eccentric (henceforth called the detection problem) and (ii) classify the events as non-eccentric $(e = 0)$, moderately eccentric $(e \in (0, 0.2])$, and highly eccentric $(e \in (0.2, 0.5])$. On the detection problem, our best performing network  detects eccentricity with $0.914$ accuracy and true and false positive rates of $0.862$ and $0.138$, respectively. On the classification problem, the best performing network classifies signals with $0.853$ accuracy. We find that our trained detector displays close to ideal behavior for the data we consider.
\end{abstract}

%Furthermore, confusion matrices, , and NN-Output scatter plots indicate that our trained network as a detector and a classifier is very close to ideal behavior. In this work, we think that testing our network on real catalogue events is out-of-scope, but is certainly an avenue for future work. Specifically, since we trained on synthetic gaussian noise using O4 representative \texttt{PSD}s, our current network will certainly not be optimal for real events. Additionally, the real events would also require the training data to include spin, which hasn’t been included in this work.
%(henceforth called the classification problem).
%(defined as the one displaying lowest testing loss) at the default threshold value

\maketitle

\section{INTRODUCTION}

The LIGO-Virgo-KAGRA network~\cite{aLIGO, Virgo, KAGRA:2020tym} of ground-based interferometric detectors has completed three observing runs. These have recorded $\mathcal{O}(100)$ gravitational wave (GW) signals, all of which were produced by compact binary coalescences (CBCs)~\cite{gwtc-1, gwtc-2, abbott2021gwtc, ias-1, ias-2, ias-3, ias-4, Nitz:2021zwj}. While the majority of these are binary black hole (BBH) mergers, binary neutron stars (BNS)~\cite{GW170817-detection, GW190425-detection} and neutron-star black-hole (NSBH) binaries~\cite{NSBH-Discovery} have also been observed.  Bayesian parameter estimation analyses (see, e.g:~\cite{Cutler:1994ys}) of these signals have revealed some of them to have interesting source parameter values that could have important astrophysical implications. These include binaries with component masses that may lie in the upper and lower mass-gaps~\cite{LIGOScientific:2020iuh, LIGOScientific:2020zkf}, signals where subdominant modes of GW radiation have been observed (see, e.g.,~\cite{LIGOScientific:2020stg}), binaries with anti-aligned and precessing spins~\cite{abbott2021gwtc, Hannam:2021pit}, and one signal that may contain 
non-standard signatures including (but not restricted to) dynamical captures ~\cite{Gamba:2021gap}, or (speculatively) even exotic compact objects~\cite{Bustillo:2020syj}.

Compact-object binaries evolved from stellar binaries radiate away orbital eccentricity early during their inspiral and circularize their orbits~\cite{Blanchet:2013haa}. This is why most of the merger events recorded so far have not 
conclusively exhibited signatures of residual eccentricity (with the notable exception of GW190521~\cite{LIGOScientific:2020ufj,Romero-Shaw:2020thy} and a few other events~\cite{Romero-Shaw:2022xko, Romero-Shaw:2021ual}). 
There is, however, a prominent class of binary black holes that are formed in dense stellar environments, such as core-collapsed globular clusters or galactic nuclei, that are expected to enter the frequency band of ground-based GW detectors with non-negligible eccentricity \cite{Sippel_2012,Strader_2012,arxiv180208654,Samsing_2017,Samsing_2014,Samsing_2018,Samsing:2017xod,Samsing:2017plz,Samsing:2016bqm,Randall:2017jop,Samsing:2017jnz,Samsing:2017oij,Leigh:2017wff,Antonini:2015zsa,Gondan:2017wzd,Antonini:2018auk,Huerta:2014eca,Chen:2020lzc,Takacs:2017wnn,Arca-Sedda:2018qgq,Antonini:2016gqe,Gondan:2018khr,Randall:2018qna,Hoang:2017fvh,Lopez:2018nkj,Rodriguez:2018pss,Zevin:2018kzq}. 
Studies from just the first two observing runs of the LIGO-Virgo detectors have suggested that we might expect merger rates for eccentric BBHs up to $100\mathrm{Gpc}^{-3}\mathrm{yr}^{-1}$~\cite{LIGOScientific:2019dag} and for eccentric BNS mergers up to $1700\mathrm{Gpc}^{-3}\mathrm{yr}^{-1}$~\cite{Nitz:2019spj}. While analyses with recent data would revise these limits, future observation runs are expected to progressively reduce the lower limit of the frequency band of the LIGO-Virgo detectors~\cite{KAGRA:2013rdx, Hall:2019xmm}, increasing the probability of detecting eccentric signals in O4, O5 and beyond (see, e.g.,~\cite{mapelli2020} and references therein).

A robust way to constrain this eccentricity is via Bayesian parameter estimation (PE), where one samples the GW likelihood/posterior for each individual event. While doable in principle, in practice such an approach has two non-trivial hurdles. The first is that currently, there do not exist any robust GW waveform models that adequately span the entire range of possible eccentricity values.
This necessitates an un-physical cutoff on the prior outside the regime of validity, thus preventing the sampling of the likelihood in that range.
The second is computational cost. Bayesian PE, especially involving eccentricity as a free parameter in addition to the standard source-parameters, is expensive both in terms of wall-clock time and number of CPU cycles required. This is a direct result of waveform approximants being slow to evaluate, and challenges the practicability of sampling the posterior of each and every detected event ($\mathcal{O}(100)$ in O4 and $\mathcal{O}(1000)$ in O5).

\begin{figure}[ht!]
\centering
\includegraphics{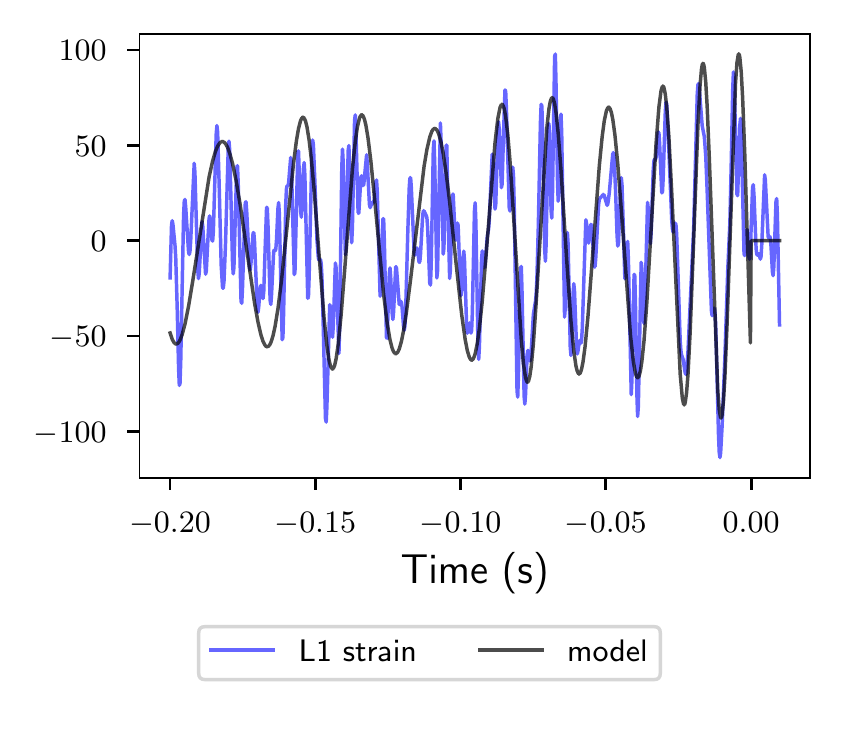}
\includegraphics{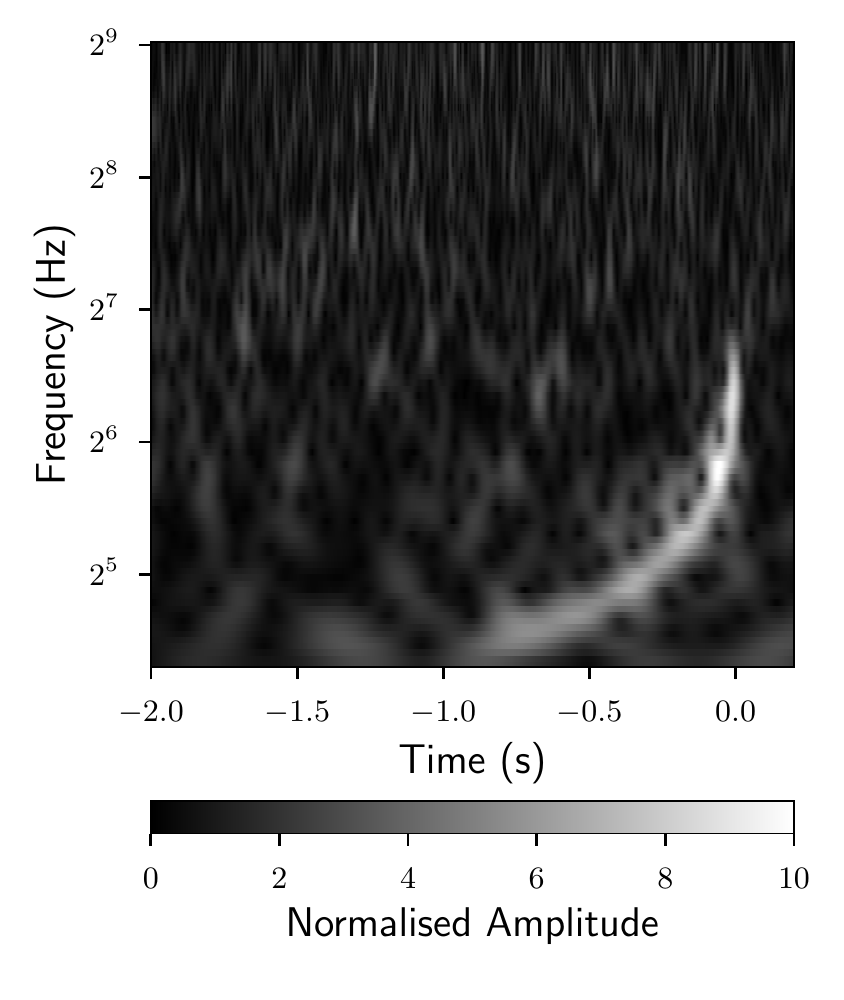}
\caption{Whitened Strain in Livingston (L1) Detector with corresponding Q-scan. (Top) Plot of waveform generated using the model \texttt{EccentricTD}, with the parameters $m_1=20.1 \ M_\odot$, $m_2=28.3 \ M_\odot$, $\mathrm{SNR}=30$, $e=0.15$.
This waveform was embedded in Gaussian noise drawn from a LIGO-O4 representative \texttt{PSD} \cite{KAGRA:2013rdx}, and then whitened with that PSD. (Bottom) The corresponding Q-transform spectrogram, where the space is divided into logarithmically spaced time-frequency tiles, and each tile has a pixel value that encodes the power of that tile. These Q-transforms are used as input images for training our neural network.}
\label{fig:L1 wf and qs}
\end{figure}

\begin{figure*}[ht!]
\includegraphics[scale=1]{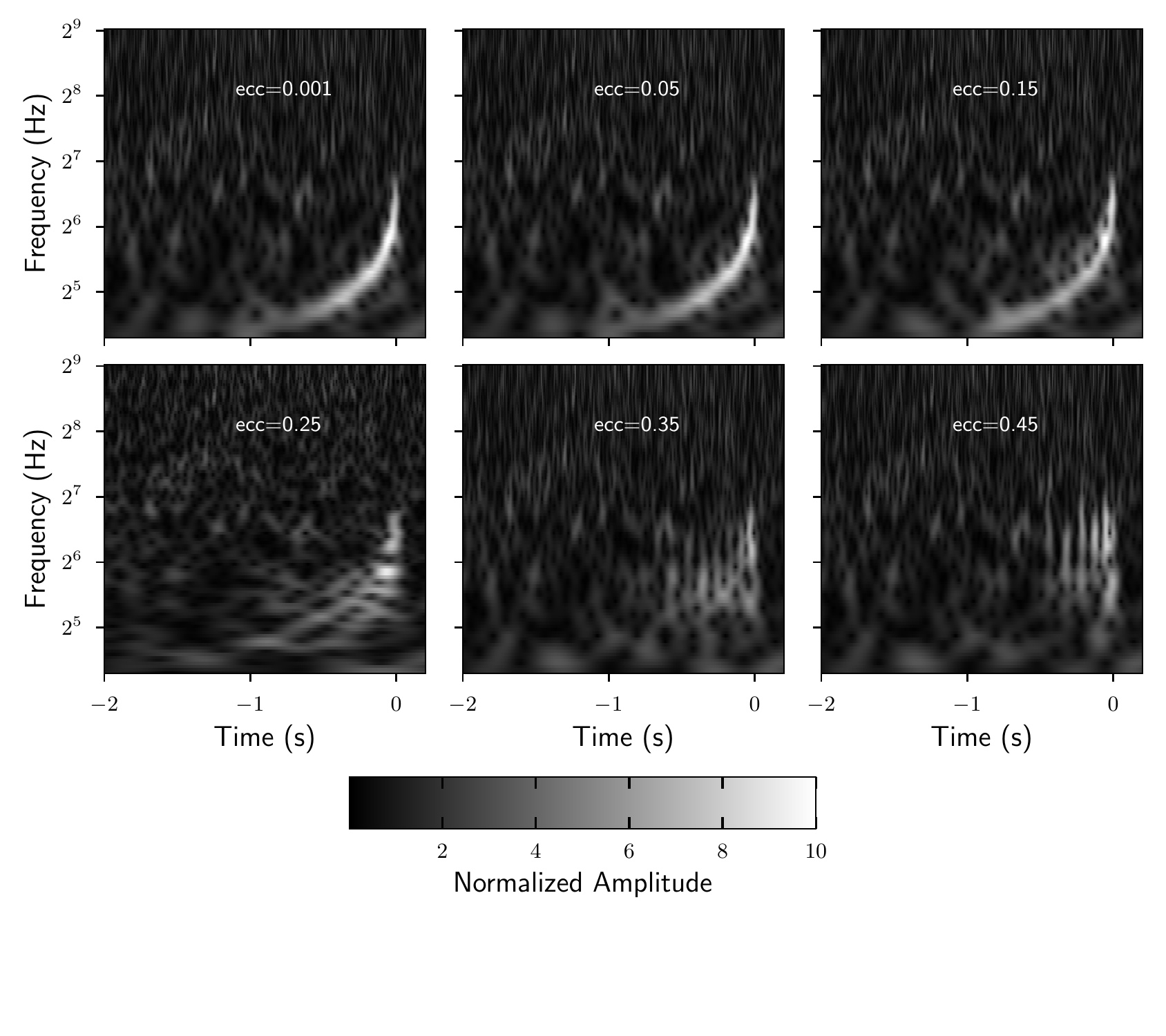}
\caption{Q-scans are plotted by varying eccentricity and keeping all other parameters fixed (to the values {$m_1=20.1M_\odot$, $m_2=28.3M_\odot$, SNR=30}) in Figure~\ref{fig:L1 wf and qs}. There is a visually observable modulation of the Q-scans as one changes the eccentricity, and hence the use of a Deep Neural Network for classification of these samples is justified.}
% Each image here is a $256\times256\times3$ image, where the each channel corresponds to the Q-scan of a different detector. More precisely, the convention we use is that the R, G and B channels correspond to the LIGO-Hanford (H1), LIGO-Livingston (L1) and Virgo (V1) detectors respectively.
\label{fig:Q-scans-with-ecc}
\end{figure*}

In order to address the second hurdle, we propose a scheme involving machine learning (ML) that could rapidly rule out the majority of events as non-eccentric, leaving the remaining to be further analysed with PE. As a proof of principle, we present a neural network architecture, primarily using separable convolutional layers~\citep{chollet2017xception}, to detect eccentricity, and also separately classify eccentricity into non-eccentric ($e=0$), moderately-eccentric ($e \in (0,0.2]$), and highly-eccentric ($e \in (0.2,0.5]$). We use time-frequency representations of the detector strain, called Q-transform spectrograms (henceforth also called Q-scans in the paper)~\cite{Chatterji:2004qg}, as our primary data representation (See Figure~\ref{fig:L1 wf and qs} for an illustrative example.)  
Broadly, Q-transforms divide the time-frequency space into bins, and record the signal energy in each tile. The expectation is that eccentricity modulates the shape of these Q-transform patterns in a detectable way (see Figure~\ref{fig:Q-scans-with-ecc}).

We train a separable-convolutional (based) neural network (SCNN) with spectrograms of synthetic GWs added to Gaussian noise (characterized by the O4 representative noise power spectral density (PSD)) to (i) segregate candidates as either non-eccentric or eccentric, which is henceforth called the \textit{detection problem} (ii) classify the events as non-eccentric $(e = 0)$, moderately eccentric $(e \in (0, 0.2]))$ and highly eccentric $(e \in (0.2, 0.5])$, which is henceforth called the \textit{classification problem}.

We find that on the detection problem, our best performing network, at the default threshold value, detects eccentricity with $0.914$ accuracy and true and false positive rates of $0.862$ and $0.138$ respectively. For the classification problem, the best performing network classifies signals with $0.853$ accuracy. 
We also find that after sufficient training, the receiver operating characteristic curves for both problems depict behaviors close to an ideal classifier, with area under curve value above $0.95$ consistently.

The rest of this paper is organized as follows: Section~\ref{s1:methodologies} summarizes the numerical methods we adapt in our approach, Section~\ref{results} contains results on the detection and classification performance of the neural network architecture used, Section~\ref{discussion} contains a summary of our results and some avenues for future extensions.

\section{METHODOLOGIES}\label{s1:methodologies}

\subsection{Data Generation}
Following Ref.~\citep{gebhard2019convolutional}, and employing \texttt{PyCBC} \citep{alex_nitz_2022_6646669}, we generate the training data for the neural network. We use the \texttt{EccentricTD} \citep{tanay2016frequency} approximant, as implemented in \texttt{LALSuite}~\cite{lalsuite}, to produce templates. For the detection problem, the eccentricity was uniformly sampled from $[0.01, 0.5]$ ($[0.001, 0.01]$) for eccentric (non-eccentric) samples\footnote{We make this choice because the \texttt{EccentricTD} model has pathologies at $e=0$.}. Similarly for the classification problem, the eccentricity was uniformly sampled from $[0.01, 0.2]$ ($[0.2, 0.5]$) for moderately-eccentric (\texttt{mod-ecc}) (significantly-eccentric (\texttt{sig-ecc})) samples. For non-eccentric (\texttt{non-ecc}) samples, the eccentricity was sampled as before.

 The masses of the binaries were individually sampled from a uniform random distribution in the $10-40M_\odot$ range. We only consider non-spinning black holes in this work, and leave the treatment of black holes with spin spin and orbital precession to future work. The sky locations were sampled uniformly over a two-sphere. The coalescence phase was uniformly sampled, and the inclination was sampled from a {cosine} distribution. The luminosity distance to the sources were sampled implicitly by sampling the three-detector signal-to-noise ratio (SNR) over the range $15-85$. The three-detector SNR (also referred to as the injection-SNR thorughout the paper) is $\sqrt{\mathrm{SNR}^2_{L} + \mathrm{SNR}^2_{H} + \mathrm{SNR}^2_{V}}$, where the individual terms under the square-root correspond to the SNR value of the signal detected by the Livington, Hanford and Virgo detectors respectively. This injection SNR ($\rho$) was sampled from the `universal' distribution, $p(\rho)\propto{}\rho^{-4}$ distribution~\cite{Schutz:2011tw} (tagged as \texttt{real}), appropriate for volumetrically distributed astrophysical sources. We also draw from a $p(\rho) = \mathrm{constant}$ distribution (tagged as \texttt{uniform}).

The waveform length was fixed to $16$ seconds at a sampling rate of $4096$ Hz. A $32$ second long (gaussian) noise sample was generated using the zero-detuning high-power design sensitivity curve for LIGO detectors \citep{LIGO-T1800044}, at the same sample rate. The signal was then embedded into the noise timeseries after tapering the signal, with the merger time fixed at $t=0$. An $8$ second long slice was extracted from this, $-6$ to $2$ seconds around the merger time. The signal was then bandpass-filtered between $20$ Hz and $512$ Hz and whitened. The Q-transform was constructed from this signal, the temporal range for which was set to $t \in [-2, 0.2]$ seconds. This range was chosen as it contained most of the visible chirp pattern for gravitational-wave signals within the parameter space that we were working with, while keeping the computational cost of training our neural network in check. (The computational cost we refer to here is the number of addition and multiplication operations in the input layer of the neural network, where an image with larger dimension will have more such operations.) Finally, the image was cropped to fix the image size to $256 \times 256 \times 3$ (where the third index/dimension corresponds to the number of detectors, which in our case is $3$).

\subsection{Neural Network Architecture}

\begin{figure*}[ht!]
\centering
\includegraphics[scale=0.68]{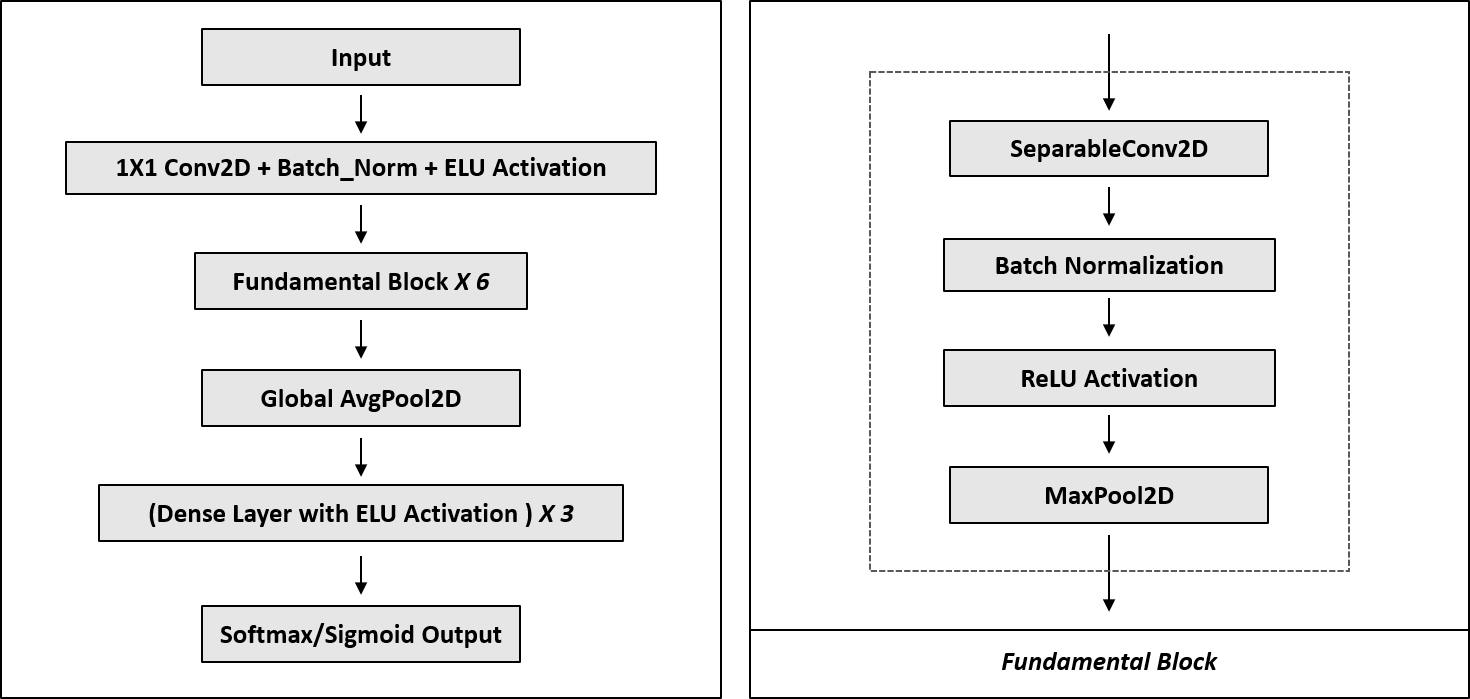}
\caption{The \texttt{ecc\_sepcnn} NN Architecture. This architecture takes in a $256\times 256 \times 3$ image data as the input, where the $3$ channels in the image correspond to the Q-scans of the three detectors. The architecture consists of a $1\times1$ convolutional block, followed by $6$ iterations of the fundamental block, followed by dense layers. These lead into a $1$ unit \texttt{Sigmoid} output for the detection problem, and a $3$ unit \texttt{Softmax} output for the classification problem. The final model was $1.2$ MB in size, and had $85,559$ trainable parameters.}
\label{fig:ecc_sepcnn NN arch}
\end{figure*}

Deep learning allows computational models that are composed of multiple processing layers to learn representations of data with multiple levels of abstraction \citep{deeplearningnature}. Image classification is one of the major applications of deep learning eg. classification on the ImageNet dataset through Deep Convolutional Neural Networks \citep{10.1145/3065386}. It can also be shown that a Deep Convolutional Neural Network (CNN) is universal, in that it can be used to approximate any continuous function to an arbitrary accuracy when the depth of the neural network is large enough \citep{https://doi.org/10.48550/arxiv.1805.10769}. Deep learning techniques are  proving to be of crucial importance in astronomy research. In the context of GW data analysis, there have been multiple works that use deep learning and other learning techniques to detect GW sources (eg.~\cite{George:2017pmj,Gabbard:2017lja, Schafer:2022dxv, Jadhav:2020oyt, Chatterjee:2022ggk}, estimate source parameters (eg.~\cite{Gabbard:2019rde, Green:2020dnx, Green:2020hst, Dax:2021tsq}), and also identify and alleviate noise artifacts in data (eg.~\cite{Zevin:2016qwy}).

Our model architecture \texttt{ecc\_sepcnn}, as depicted in Figure~\ref{fig:ecc_sepcnn NN arch}, primarily uses \texttt{Separable-Convolutional} (\texttt{SeparableConv2d}) layers instead of the usual \texttt{Convolutional} layers. A (depthwise-) separable convolution consists of a depthwise convolution, performed independently over each channel of an input, followed by a pointwise $1\times1-$convolution. The former consists of using a separate convolutional kernel for each input channel and the latter consists of linearly combining all pixel data with linear coefficients being learn-able parameters. This is different to the normal convolution, where each kernel has depth equal to the number of input channels. The mixing of channels is done in the same step as the convolution operation. In a sense, the depthwise-separable convolution factorizes the task of a regular convolution, and 
this factorization has the effect of drastically reducing computation and model size while not sacrificing much on accuracy (for details on the degree of reduction, refer to \cite{DBLP:journals/corr/HowardZCKWWAA17}). These layers are known to efficiently extract cross channel correlations, with fewer parameters when compared to regular \texttt{Convolutional} (\texttt{Conv2D}) layers \citep{chollet2017xception}. In this architecture, as one moves through the network, the number of channels in layers continues to increase while the height and width of the layer decrease.

All convolutional layers are followed by batch normalization layers. \texttt{Batch Normalization} is known to have a regularization-like effect by keeping the input to the next layer close to $0$, like the \texttt{L1} and \texttt{L2} regularization schemes \cite{ioffe2015batch}. The activation function chosen for the \texttt{Convolutional} layers was \texttt{ReLU}, whereas the activation for \texttt{Dense} layers was \texttt{ELU}, since the latter has been known to work better for image classification tasks on the \texttt{ImageNet} dataset. \citep{mishkin2017systematic} 

The input layer takes in the $256 \times 256 \times 3$ image, where $3$ channels correspond to LIGO-Hanford (H1), LIGO-Livingston (L1) and Virgo (V1) detectors. This layer will need to be modified to a greater number of channels if data from more or different detectors needs to be added to the input in the future.

The first part consists of a convolutional block, which consists of the following. A convolutional layer ($16$ filters with $1 \times 1$ kernels) followed batch normalization and an exponential linear unit (\texttt{ELU}) layer. The intuition behind this was to linearly combine the pixel data from each detector, with the linear coefficients being learn-able parameters.
The next part contains separable-convolutional blocks, which consist of the following. A \texttt{SeparableConv2D} layer followed by \texttt{Batch Normalization}, Rectified Linear Unit (\texttt{RelU}) activation, and a \texttt{Max Pooling} layer. The filter sizes in the \texttt{SeparableConv2D} go from $5\times5 \rightarrow 5\times5 \rightarrow 5\times5 \rightarrow 3\times3 \rightarrow 3\times3$, where the intuition of moving from bigger filters to smaller filters is drawn from \texttt{AlexNet}\citep{NIPS2012_c399862d}. In the same, the channel numbers go from $32 \rightarrow 64 \rightarrow 128 \rightarrow 128 \rightarrow 256$.
To connect the \texttt{SeparableConv2D} to \texttt{Dense} layers, a \texttt{Global Average Pooling} is implemented. $3$ \texttt{Dense} layers are connected, with decreasing number of nodes in the network. \texttt{Dropout} is implemented before the dense layer with high number of parameters to prevent any potential overfitting. Also, each convolutional and separable convolutional layer has \texttt{L2} kernel and bias regularizers implemented to reduce overfitting.

The output layer is a $1$ unit \texttt{Softmax} for the eccentricity detection problem, where $1$ corresponds to an eccentric signal (eccentricity $\in (0.001,0.5)$) and $0$ to a non-eccentric signal. \texttt{Binary Cross-Entropy} is used as the error function, and \texttt{Adam's Optimizer} \citep{kingma2014adam} was used with an initial learning rate of $5$ $\times$ $10^{-4}$.
For the eccentricity classification problem, the output layer is a $3$ unit \texttt{Softmax}, where the labels are the one-hot-encodings for the $3$ classes. These $3$ classes are non-eccentric signals (label $=\ [1,0,0]$), moderately eccentric signals with eccentricity $\in (0.001,0.2)$ (label $=\ [0,1,0]$), and significantly eccentric signals with eccentricity $\in (0.2,0.5)$ (label $=\ [0,0,1]$). \texttt{Categorical Cross-Entropy} is used as the error function, and \texttt{Adam's Optimizer} \citep{kingma2014adam} was used with an initial learning rate of $5 \times 10^{-4}$.

The final model was $1.2$ MB in size, and had $86,807$ parameters 
,out of which $85,559$ were trainable. The block diagram representation of this network is given in Figure \ref{fig:ecc_sepcnn NN arch} .

\section{RESULTS}\label{results}
\begin{figure}[ht!]
    \centering
    \includegraphics{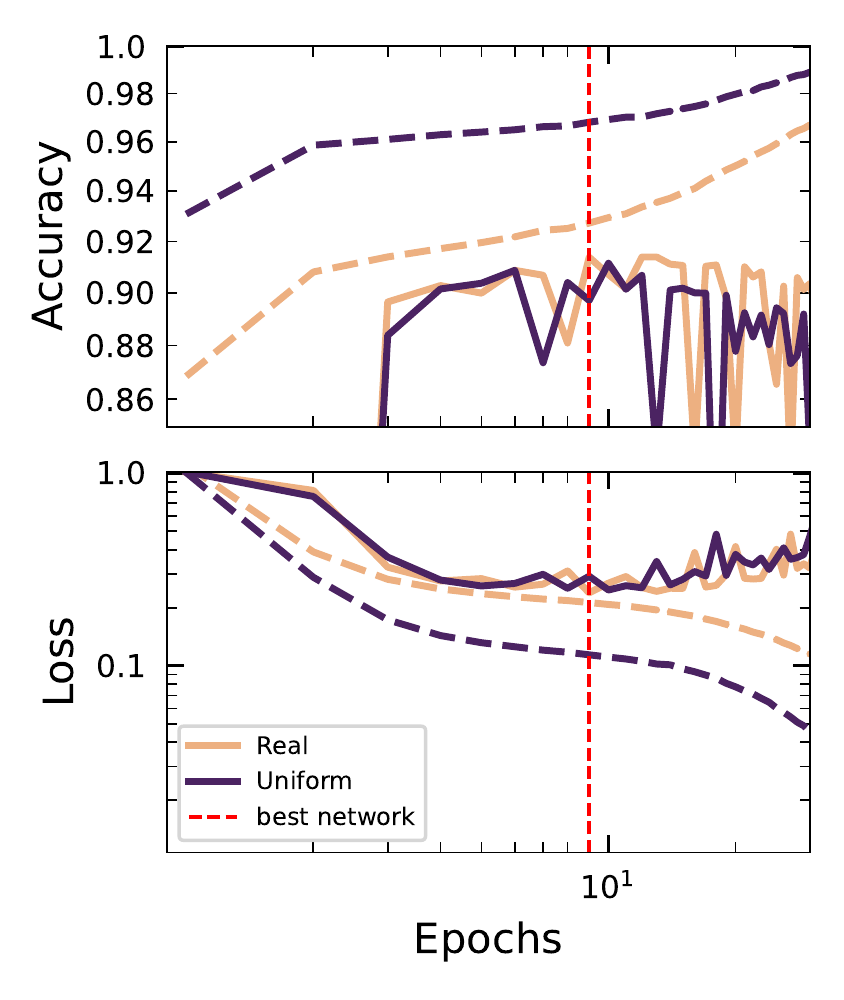}
    \caption{Accuracy and Loss Plots for the detection problem (dashed and solid lines respectively correspond to training and testing accuracy/loss). The training accuracies for the real distribution crosses $\sim .92$ around $9$ epochs, which showed the least testing loss and was defined as the best performing NN (red dashed line). The testing accuracy at this best performing NN was $0.914$. We observed better performance (lower testing loss and higher testing accuracy) for \texttt{real} training in general. We also observed the onset of overfitting around the $12^{\rm th}$ epoch with of divergence in the training and testing lines (for a given training set distribution), which is why we stop training at $30$ epochs.}
    \label{fig:bin_loss_acc}
\end{figure}
\begin{figure}[ht!]
    \centering    
    \includegraphics{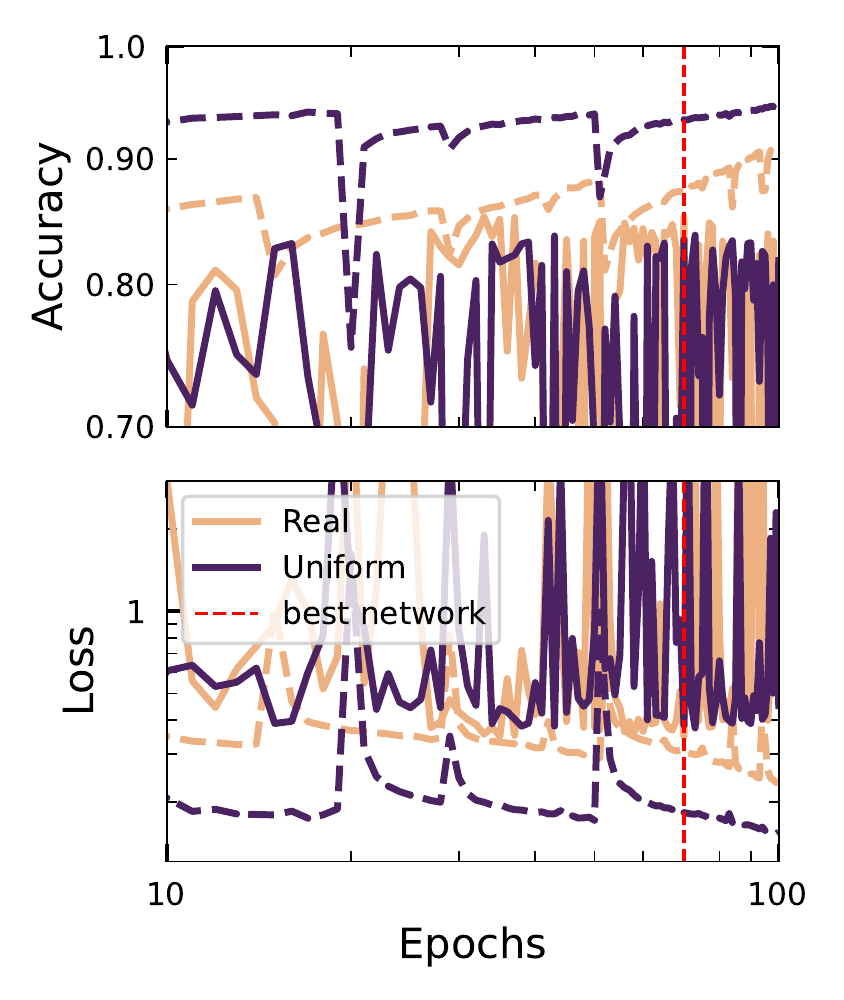}
        \caption{
        Accuracy and Loss Plots for the classification problem (dashed and solid lines respectively correspond to training and testing accuracy/loss). The training accuracies for \texttt{real} and \texttt{uniform} training set distribution respectively, rose to $\sim 0.88$ around $70$ epochs. The network at $70$ epochs of \texttt{real} training showed the least testing loss, which was defined as the best performing NN (red dashed line). The testing accuracy at this best performing NN was $0.853$. We again observed better performance (lower testing loss and higher testing accuracy) for \texttt{real} training. In this case we don't clearly observe the onset of overfitting, but observe rapidly oscillating accracies and losses starting at the $40^{\rm th}$ epoch. Since accuracy doesn't improve beyond the $70^{\rm th}$, we stop training at $100$ epochs.}
    \label{fig:tri_loss_acc}
\end{figure}
\subsection{Training and Testing}
The training and testing processes were implemented via the machinery of \texttt{tensorflow.keras} \cite{tensorflow2015-whitepaper}, on $48$ 
cores of an AMD EPYC 7352 processor. 
The training sets for both problems contained $10^5$ samples of $256\times256\times3$ images, where the total samples were divided equally between the $2$ (binary/detection problem) or $3$ (ternary/classification problem) classes. As previously mentioned, the training sets were labelled \texttt{real} for indicating a $\rho^{-4}$ injection SNR distribution; and \texttt{uniform} for a uniform injection SNR distribution. The batch size was fixed to $500$, and the Adam Optimizer was implemented with an initial learning rate of $5 \times 10^{-4}$. Training was carried out for $30$ epochs for the detection problem and $100$ epochs for the classification, and the training accuracies converged to $>.92$ and $~0.90$ respectively within the given epochs. The networks were also separately trained on \texttt{uniform} and \texttt{real} sets to observe which yields better testing accuracies. 

For the detection problem, the accuracies were evaluated with the default threshold value of $0.5$, meaning that for an input sample, if the sigmoid output of the NN was greater than $0.5$, it was assigned the eccentric-class and for values smaller than $0.5$, the non-eccentric class was assigned. For the classification problem, each sample was assigned the class that corresponded to the softmax node with the largest output.
The test sets contained $2\times 10^4$ and $3\times 10^4$ samples for the detection and classification problems, where the batch size was again fixed to $500$ for both. The testing was only done on \texttt{real} sets, because one expects the set of observed GW signals to follow the $\rho^{-4}$ distribution. It was observed that \texttt{real} training produced better performing networks for both problems, in that the peak testing accuracy is better and the convergence to the peak is faster.

The training time for the detection problem was $8$ seconds per batch, while that for the classification problem was $11$ seconds per batch. The testing time for both problems was observed to be $0.82-2$ ms per batch. 
The training times are different because the loss functions for the two problems are different; the detection problem uses ``binary-cross-entropy'', whereas the classification problem uses  ``categorical-cross-entropy''. The latter involves a greater number of arithmetic operations, hence leading to greater training time due to a more involved back-propagation step. The testing times are roughly the same because the network architectures except the output layer are exactly the same, and time difference due to the higher number of operations in the output layer for the classification problem while forward-propagation is not significant.

For the detection problem, the training accuracies, for the both distributions converged to $>0.92$ around $9$ epochs.  The network at $9$ epochs of \texttt{real} training showed the least testing loss (which was defined as the best performing NN). The testing accuracy at this best performing NN was $0.914$. We observe better performance for \texttt{real} training, where better performance is defined as having lower testing loss and higher testing accuracy. Though the uniform-distribution training accuracy seems to have higher values in Figure-\ref{fig:bin_loss_acc}, the testing accuracy(/loss) is lower(/higher), which is what is used as the metric of performance.

For the classification problem, the training accuracies, for \texttt{real} and \texttt{uniform} training set distribution respectively, rose to $\sim 0.9$ around $20$ epochs. The network at $70$ epochs of \textbf{real} training showed the least testing loss (which was defined as the best performing NN). The testing accuracy at this best performing NN was $0.853$. We again observe better performance (lower testing loss and higher testing accuracy) for \texttt{real} training. The \texttt{uniform} training converges to a higher accuracy and lower loss while training, but the testing lines are observed to be worse (higher testing loss and lower testing accuracy). We also observe rapid oscillations starting at the $40^{\rm th}$ epoch without any improvement in performance, which is why we stop training at $100$ epochs. Again, though the uniform-distribution training accuracy seems to have higher values in Figure-\ref{fig:bin_loss_acc}, the testing accuracy(loss) is lower(higher), and hence the real-trained network again performs better.

\subsection{Testing the networks with \texttt{TEOBResumS}}
\label{app:teobresums}
To test whether the detection and classification learned by our model was specific to the waveform model it was trained on, or whether the patterns learned were general to other eccentric waveform models as well, we generated a test set with a different waveform model, \texttt{TEOBResumS}~\cite{Riemenschneider:2021ppj, Nagar:2020pcj, Nagar:2019wds, Nagar:2018zoe, Nagar:2015xqa, Damour:2014sva}, and ran it through our trained networks. The reason behind running this test set through all trained networks as opposed to just the best performing one was to show that the best network didn’t just happen to give out a good accuracy value at the best EccentricTD trained epoch, but that the accuracy converged to a high value around the best trained epoch.

\begin{figure}[ht!]
    \centering
    \includegraphics{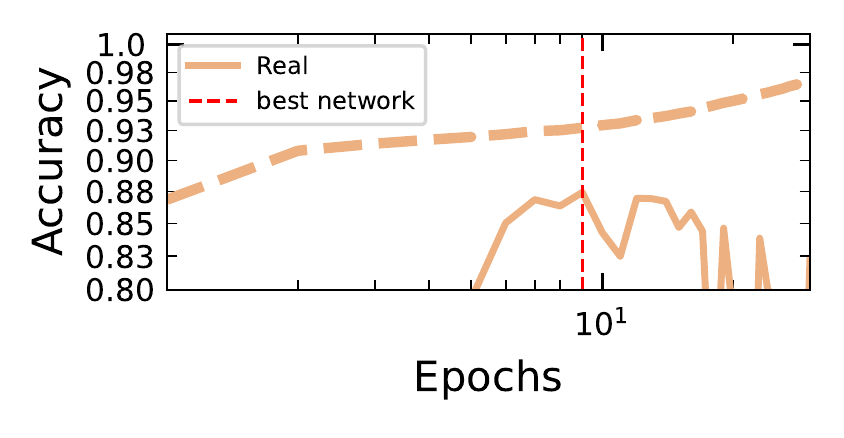}
    \caption{Accuracy Plot for the detection problem on \texttt{TEOBResumS} test-sets, dashed and solid lines respectively correspond to training (on EccTD) and testing accuracy (on \texttt{TEOBResumS}). The testing accuracy for best performing NN (\texttt{real} trained for $9$ epochs) was observed to be $0.88$.}
    \label{fig:bin_loss_acc_teob}
\end{figure}

\begin{figure}[ht!]
    \centering    
    \includegraphics{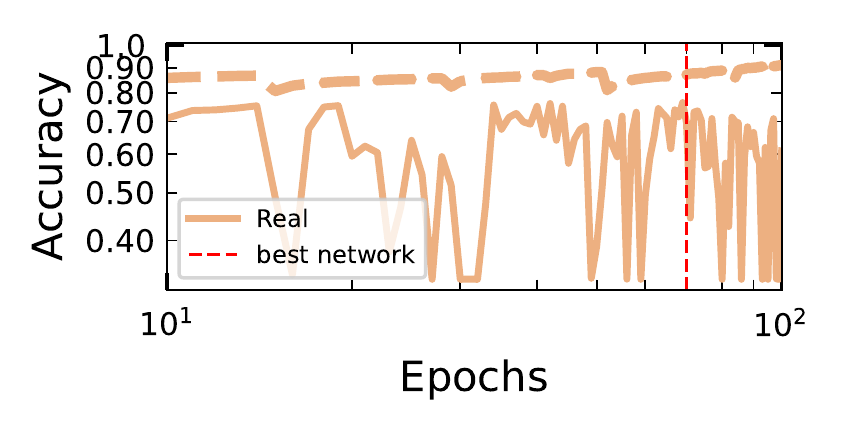}
    \caption{Accuracy Plot for the classification problem on \texttt{TEOBResumS} test-sets, dashed and solid lines respectively correspond to training (on EccTD) and testing accuracy (on \texttt{TEOBResumS}). The testing accuracy for best performing NN (\texttt{real} trained for $70$ epochs) was observed to be $0.76$.}
    \label{fig:tri_loss_acc_teob}
\end{figure}

For the detection problem, Figure~\ref{fig:bin_loss_acc_teob}, the testing accuracy for best performing NN (\texttt{real} trained for $9$ epochs) was observed to be $0.88$. For the classification problem, Figures~\ref{fig:tri_loss_acc_teob}, the testing accuracy for best performing NN (\texttt{real} trained for $70$ epochs) was observed to be $0.76$. The detection network appears to be less sensitive to the waveform model compared to the classification network, as can be seen from a $~3\%$ decrease in performance of the detector as opposed to the $~10\%$ decrease in performance of the classifier. 

For the detection problem, the slight decrease in accuracy on \texttt{TEOBResumS} waveforms could be attributed to the the presence of the full inspiral, merger and ringdown parts of the signal in the model, while EccentricTD only contains the inspiral part. However only a small accuracy decrease implies that waveform systematics do not affect network performance significantly, at least at the SNRs and detector network configurations considered in this work. 

\subsection{Confusion Matrices}

\begin{figure*}[ht!]
    \centering
    \includegraphics[width=0.95\columnwidth]{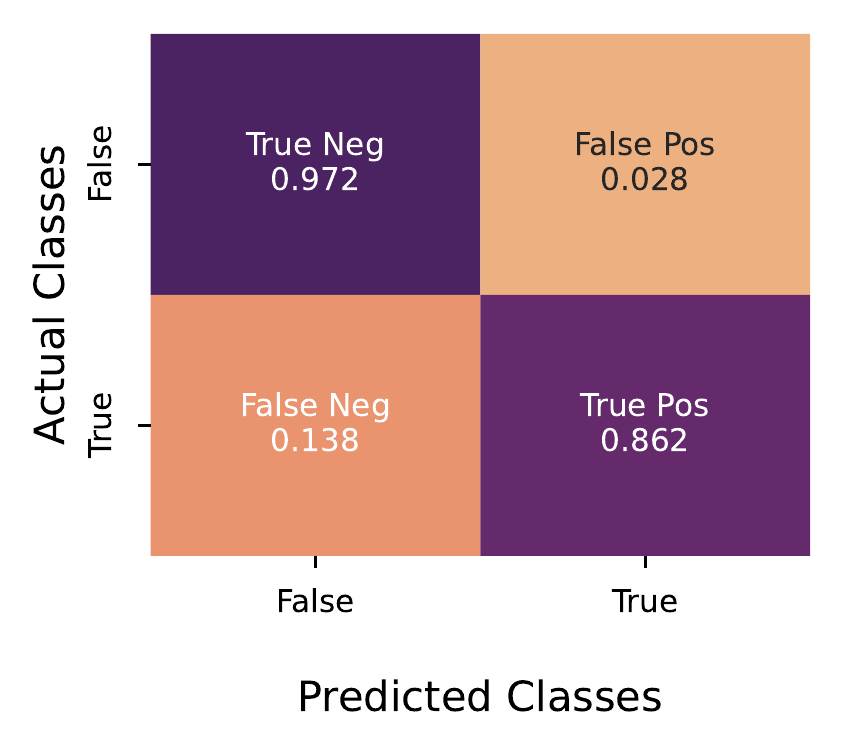}
    \includegraphics[width=0.95\columnwidth]{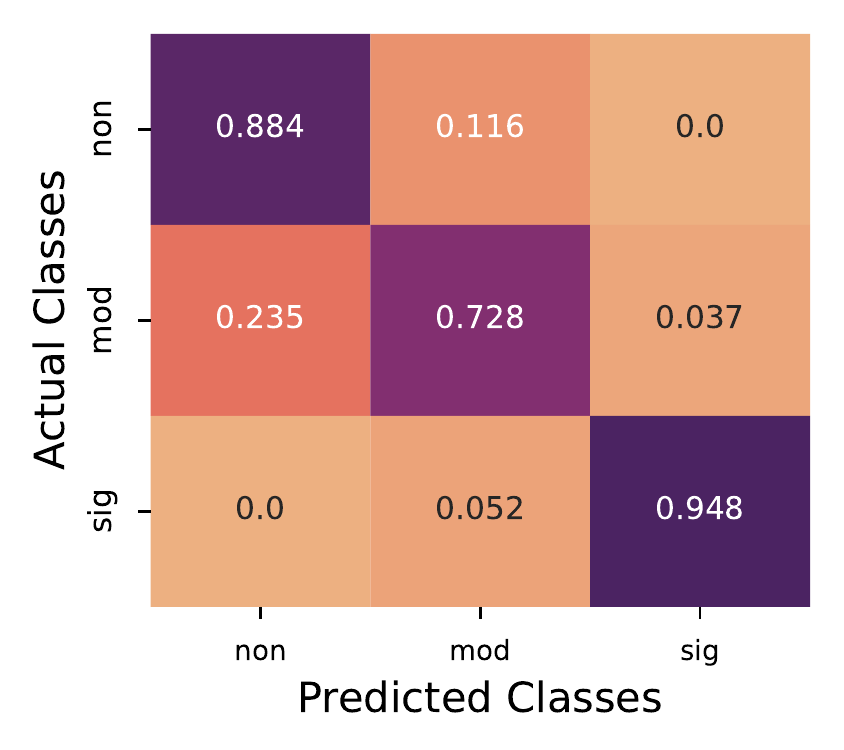}
    \caption{Confusion Matrices for the detection and classification problem. For the detection problem (left plot), the number of classes is $2$, and hence the confusion matrix for the ideal classifier is $\mathbb{1}_2$. We observe that the confusion matrix is close to an ideal classifier, with the diagonal elements $0.972$ and $0.862$ respectively.
    For the classification problem (right plot), the number of classes is $3$, and hence the confusion matrix for the ideal classifier is $\mathbb{1}_3$. We observe that the confusion matrix to be close to an ideal classifier for the \texttt{non-ecc} and \texttt{sig-ecc} classes, with the diagonal elements $>0.884$ and $>0.948$ respectively. We also observe that the true \texttt{mod-ecc} class has the most misclassifications, with $\sim 23.5\%$ and $\sim 3.7\%$ samples classified as \texttt{non-ecc} and \texttt{sig-ecc} respectively. Most signals are also misclassified to $\texttt{mod-ecc}$ predicted class.}
    \label{fig:confusion_matrix}
\end{figure*}

Confusion matrices are tools to estimate the performance of a supervised machine learning algorithm. The rows enumerate the true labels of inputs, and the columns enumerate the predicted classes. Each sample when run through a neural network conforms to a cell in such a matrix, since it has an original and known label, and a predicted label of the class that the network classifies it into. In our convention, the value in each cell of the confusion matrix is the number of samples that fall into that category normalized with the total number of samples in that row. An ideal classifier hence corresponds to a confusion matrix that is the Identity matrix of dimension equal to the number of classes in the problem ($\mathbb{1}_{n}$, where $n$ is the number of classes). 
Furthermore, for the confusion matrices for the detection problem, the numerical values in each cell also correspond to the category-rates. For example, the value in the True-True cell corresponds to the True Positive Rate (TPR) for the network on the detection problem.
We plot the confusion matrices for the best performing network (defined as the one displaying lowest testing loss), after running a set containing $2\times 10^4$ samples through the network, and collecting the true and predicted label information. The confusion matrices are plotted in Figure~\ref{fig:confusion_matrix}.

For the detection problem, the number of classes is $2$, and hence the confusion matrix for the ideal classifier is $\mathbb{1}_2$. We observe that the confusion matrix is very close to an ideal classifier, with the diagonal elements $0.972$ and $0.862$ respectively.
For the classification problem, the number of classes is $3$, and hence the confusion matrix for the ideal classifier is $\mathbb{1}_3$. We observe that the confusion matrix to be close to an ideal classifier only for the \texttt{non-ecc} and \texttt{sig-ecc} classes, with the diagonal elements $0.884$ and $0.948$ respectively. We also observe that the true \texttt{mod-ecc} class has the most misclassifications, with $\sim 23.5\%$ and $\sim 3.7\%$ samples classified as \texttt{non-ecc} and \texttt{sig-ecc} respectively. Most signals are also misclassified to $\texttt{mod-ecc}$ predicted class, with $11.6\%$ \texttt{non-ecc} samples and $5.2\%$ \texttt{sig-ecc} samples classsified as \texttt{mod-ecc}.

We note that of the small number false negatives, almost all have eccentricity near category boundaries. This means that if we do have false classification of an eccentric event as being non-eccentric, it likely will have a very small eccentricity. In a follow-up Bayesian parameter estimation analysis, estimating such a small eccentricity will have associated statistical uncertainties that will make the posterior distribution on $e$ consistent with $0$. Only if this rare mis-classification event has a sufficiently high SNR (which is rare in itself), these false negatives will lead to a loss in the astrophysical information extracted from the GW event. Furthermore, for the detection problem, the confusion matrix was computed using the default threshold value on the output of the neural network. This threshold can be treated as a tunable hyper-parameter, in that, its value can be tweaked to get a desired high True Positive Rate, at the cost of an increase in the False Positive Rate. (Figure-\ref{fig:fpr_fnr_thresh})

%False negatives are considered a fundamental block to using Neural Networks instead of traditional matched-filtering type algorithms so as to not let even one eccentric signal slip, this freedom of tuning the threshold value partially resolved this issue.

\begin{figure*}[ht!]
    \centering
    \includegraphics{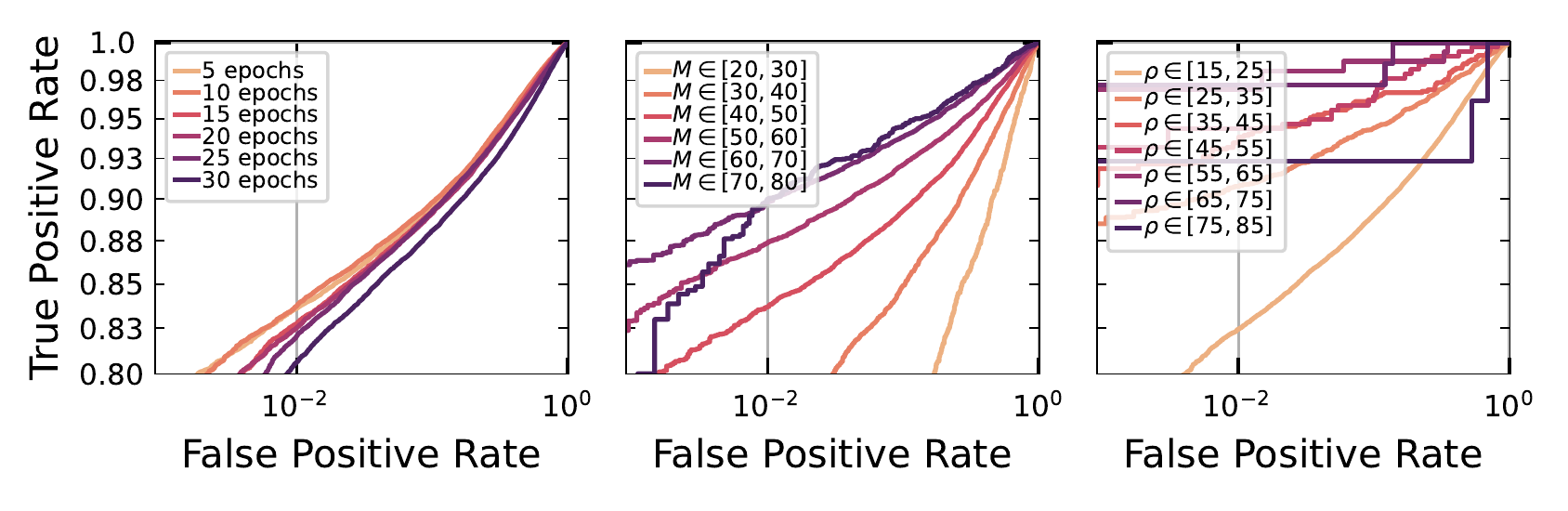}
    \caption{ROC curves for the detection problem. (Left) The observed behavior is close to an ideal binary classifier, with TPR values as high as $.8$ for FPRs as low as $.01$. The AUC values are consistently observed to be $>0.95$ for networks with greater than $9$ epochs of training, which signals proximity to ideal behavior. (Middle) The $20-30M_\odot$ and $30-40M_\odot$ slices have ROCs with lowest TPRs. This lower performance can be attributed to the signals patterns being very long, hence making the pattern in the Q-scan spread out beyond the range of our Q-scan. It can also be attributed to comparatively lower number of samples in the training set since the total mass distribution is triangular distribution in $[20,80]$. (Right) The faintest signals in the SNR range $\in [15,25]$ have the lowest TPRs. This is seen as an indication of the fact that fainter patterns are more difficult for the network to analyse and make predictions on. The next lowest TPR is in the SNR range $\in [75,85]$, which can be attributed to comparatively lower number of signals in that SNR range in a $\rho^{-4}$ distribution in the training set.}
    \label{fig:bin_ROC}
\end{figure*}

\begin{figure*}[ht!]
    \centering
    \includegraphics{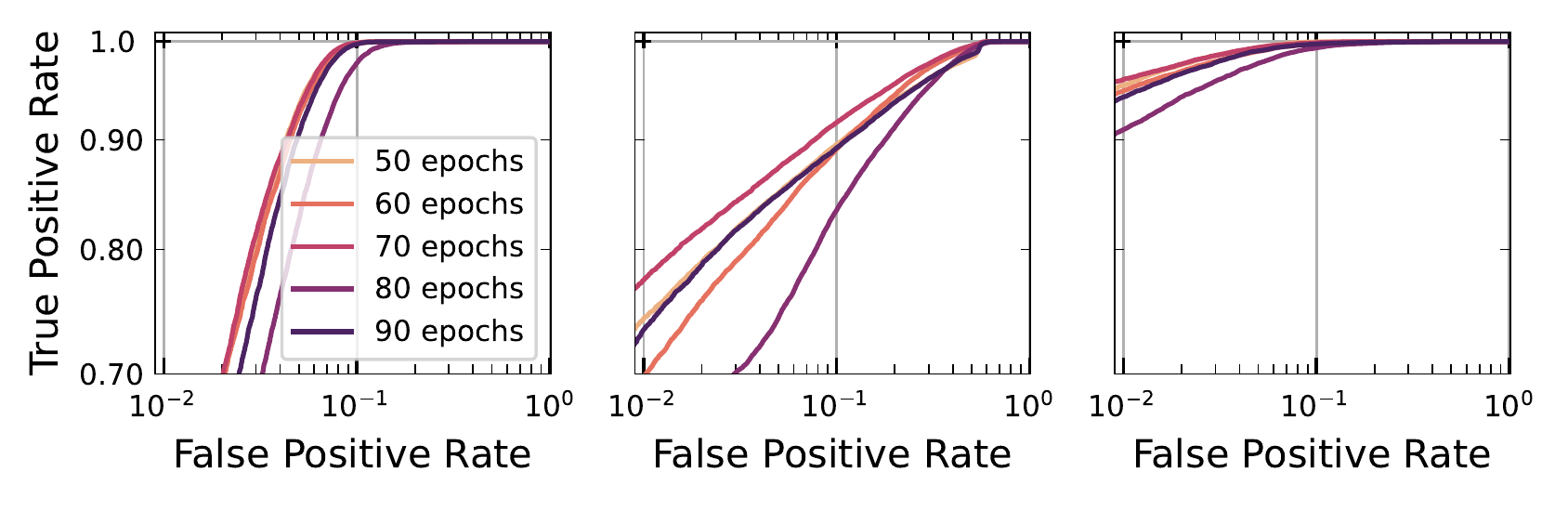}

    \caption{ROC curves for the classification problem. The \texttt{non-ecc}, and \texttt{mod-ecc} and \texttt{sig-ecc} classes were respectively considered as the positive class for the left, middle and right figures. The TPRs are observed to be higher than $.85$ for FPRs as low as $.1$. The AUC values are observed to be $>0.95$ for networks with greater than $50$ epochs of training, which again signals proximity to ideal behavior.}
    \label{fig:tri ROC}
\end{figure*}

\subsection{Receiver Operating Characteristic (ROC) Curves}
The ROC curve is a plot between the True-Positive-Rate (TPR) and False-Positive-Rate (FPR), based on the output of a machine learning algorithm on a given problem, where each point corresponds to a certain threshold value that is used to assign a class label for the value output by the algorithm \cite{FAWCETT2006861}. The ROC curve encapsulates the information about roughly how different and far apart are the NN output values for samples from different classes. For an ideal classifier, the ROC curve is just a straight line at TPR=$1$ for all FPRs. Physically, this indicates that intersection of the set of all NN output values corresponding to the positive class with the set of all NN output values corresponding to the negative class is $\phi$ or null. The Area Under Curve (AUC), as the name suggests, is just the area bounded by an ROC curve. For an ideal classifer, the AUC value is $1$.

For a binary classification problem, the ROC is simply a plot of the values in the second column of the confusion matrices generated by varying the threshold value between the the two extremes $0$ and $1$. On an ROC plot, the points $(1,1)$ and $(0,0)$ correspond to a threshold values of $0$ and $1$ respectively, with the remaining points on the curve corresponding to the threshold values in between. 
For a multi-class problem, the ROC is plotted using the ``one-vs-all'' method, where each class is separately considered the positive class and the others are considered the negative classes for the particular case. For a problem with $N$ classes, this method hence generates $N$ ROC plots.

The detection problem ROC for our best performing network (defined as the one displaying lowest testing loss) are plotted in Figure~\ref{fig:bin_ROC}. The observed behavior is close to an ideal binary classifier, with TPR values as high as $.8$ for FPRs as low as $.01$. The AUC values are consistently observed to be $>0.95$ for networks with greater than $9$ epochs of training, which signals proximity to ideal behavior. The classification problem ROCs are plotted in Figures~\ref{fig:tri ROC}. The \texttt{non-ecc}, and \texttt{mod-ecc} and \texttt{sig-ecc} classes were respectively considered as the positive class for the left, middle and right figures. The TPRs are observed to be higher than $.85$ for FPRs as low as $.1$. The AUC values are observed to be $>0.95$ for networks with greater than $50$ epochs of training, which again signals proximity to ideal behavior.

\subsection{Neural Network Output with Eccentricity and Injection SNR}
For the detection problem, the sigmoid output of the neural network is a scalar number $\in [0,1]$. Based on a threshold ($=0.5$ by default), the signals are classified as follows,
\begin{verbatim}
    if output >= threshold:
        predicted_class:= eccentric
    else:
        predicted_class:= non-eccentric
\end{verbatim}
Hence for an ideal detector, to retain an accuracy of $1$ independent of the threshold, all eccentric signals should correspond to an output=$1$ and all-non eccentric signals should correspond to output=$0$.

\begin{figure}[ht!]
\centering
\includegraphics[width=\columnwidth]{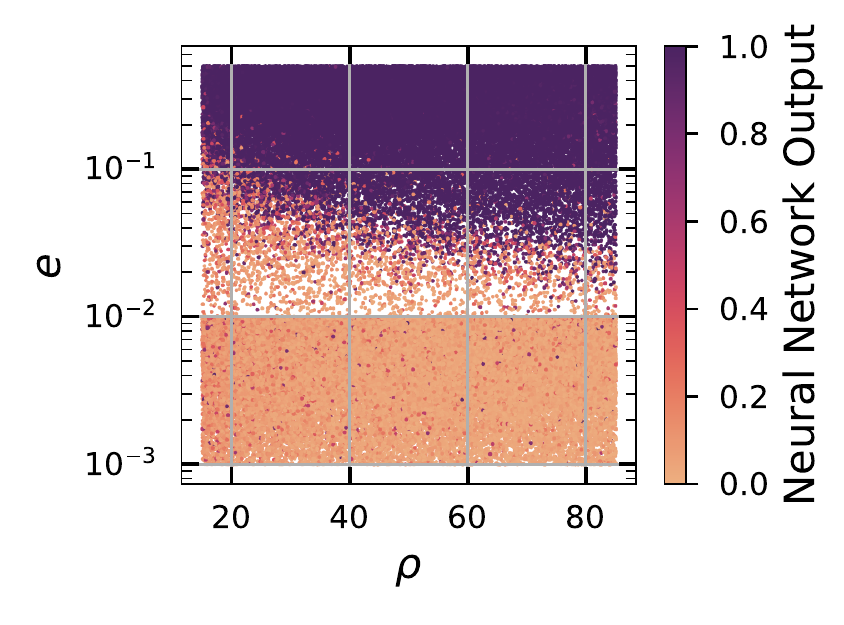}
\caption{Output of the best-performing neural network for the detection problem. Nearly all eccentric samples have an output close to $1$, and the non-eccentric samples have an output close to $0$, as was expected out of a good detector. Samples at the boundary of the eccentricity classes ($e\sim 0.01$) are the ones that are observed to have an intermediate output value. This overall behavior is also observed to be variable across SNR, with a higher number of low nn-output values leaking beyond the boundary ($e\sim 0.01$) at lower SNR.}
\label{fig:bin_out}
\end{figure}

For the classification problem, the softmax output of the neural network is a $3$-tuple ($\equiv$ out[$i$], $i=0,1,2$), with each element of the tuple $\in [0,1]$. The signals are classified as follows,
\begin{verbatim}
    if out[0] == max(out[0],out[1],out[2]):
        predicted_class:= non-eccentric
    else if out[1] == max(out[0],out[1],out[2]):
        predicted_class:= moderately-eccentric
    else:
        predicted_class:= significantly-eccentric
\end{verbatim}
Hence for an ideal classifier, all \texttt{non-ecc}, \texttt{mod-ecc} and \texttt{sig-ecc} signals should correspond to an $(1,0,0);\ (0,1,0);$ and $(0,0,1)$ respectively.

We plot a scatter-plot of the NN-output (of our best performing networks) with {eccentricity} and {injection SNR}, to observe how close our network is to the ideal detector and classifier, and to observe how the output scales with eccentricity, particularly at the boundaries of the eccentricity-classes. 

Figure~\ref{fig:bin_out} is the plot for the detection problem. Nearly all eccentric samples have an output close to $1$, and the non-eccentric samples have an output close to $0$, as was expected out of a good detector. Samples at the boundary of the eccentricity classes ($e\sim 0.01$) are the ones that are observed to have an intermediate output value. This overall behavior is also observed to be variable across SNR, with a higher number of low nn-output values leaking beyond the boundary ($e\sim 0.01$) at lower SNR 

\begin{figure*}[ht!]
\includegraphics{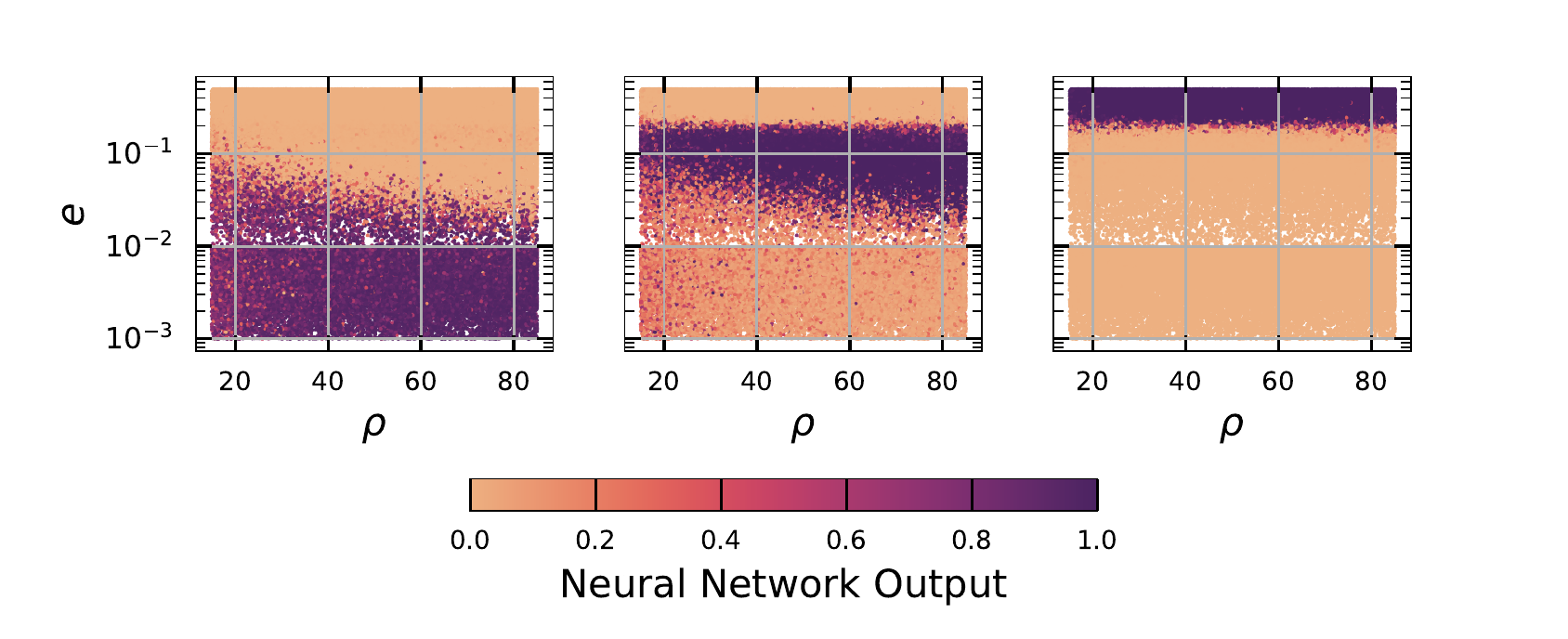}
\caption{Output of the best-performing neural network for the classification problem.
Here the first, second and third element of the output tuple is plotted with eccentricity and SNR, for the left, middle and right plots, which correspond to \texttt{non-ecc}, \texttt{mod-ecc} and \texttt{sig-ecc} respectively. For the output corresponding to the \texttt{sig-ecc} class, the values are close to $1$ for points above the \texttt{mod-sig} boundary, which is ideal. However, for the outputs corresponding to the \texttt{low-ecc} and \texttt{mod-ecc} classes, points with high values leak beyond the \texttt{non-mod} boundary, indicating that most misclassifications happen there.}
\label{fig:tri_out}
\end{figure*}

Figures~\ref{fig:tri_out} is the plot for the classification problem. For the output corresponding to the \texttt{sig-ecc} class, the values are close to $1$ for points above the \texttt{mod-sig} boundary, which is ideal. However, for the outputs corresponding to the \texttt{low-ecc} and \texttt{mod-ecc} classes, points with high values leak beyond the \texttt{non-mod} boundary, indicating that most misclassifications happen there.

\subsection{Testing Accuracies and ROC Curves for Parameter Subspaces}
For our network architectures, to test whether learning to detect and classify signals on a specific subspace of input parameters was especially difficult, we plotted the testing accuracies for different parameter sub-spaces, viz. {eccentricity}, {total mass} and injection SNR. Once again, these accuracies were evaluated at the default threshold value of $0.5$ for the detection problem. 

\begin{figure*}[ht!]
\centering
\includegraphics{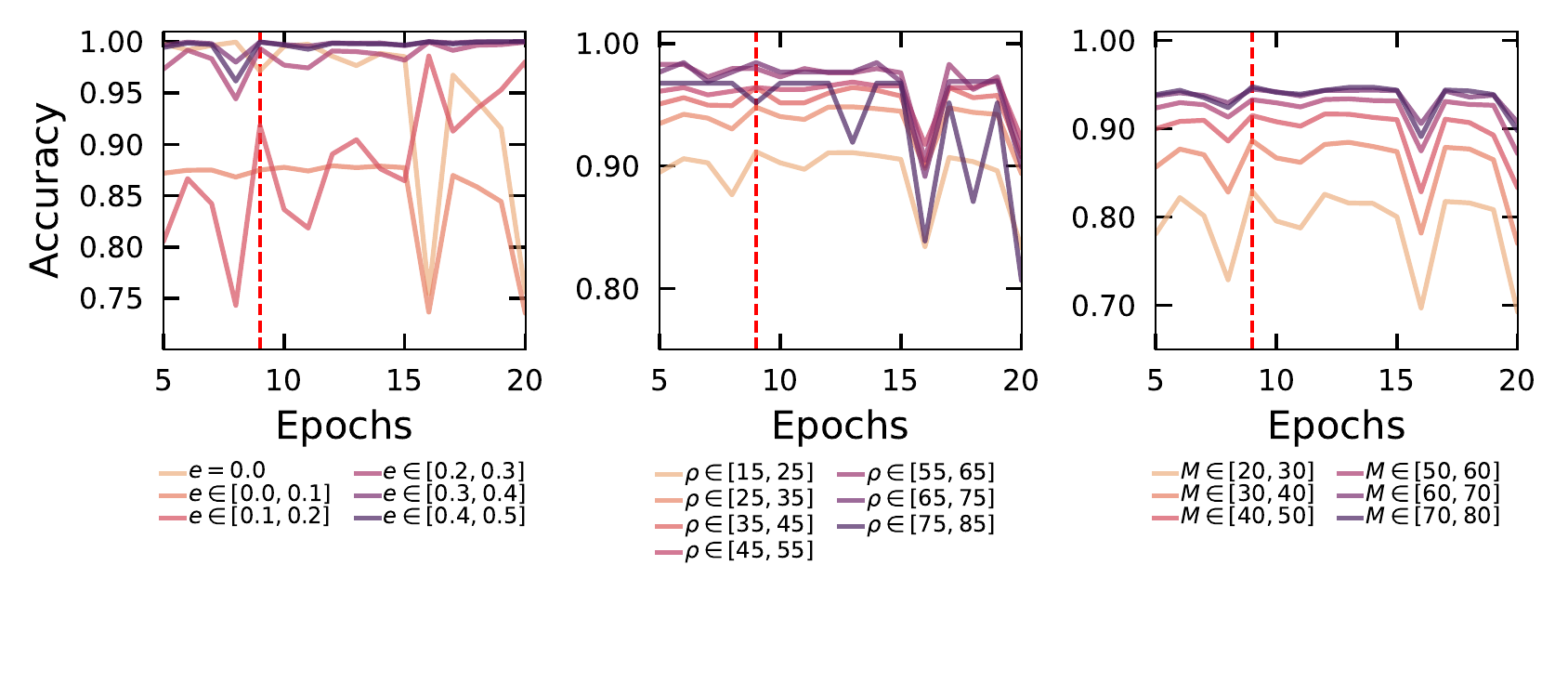}
\caption{All observations are consistent with NN output plots and slice ROC plots. The eccentricity slice $0-0.1$ has the lowest testing accuracy, since it forms the border between the two detection classes. All other slices have testing accuracies $>0.92$ after $9$ epochs. The accuracy across SNRs is almost uniform, with testing accuracies of all slices converging to $> 0.9$, with the faintest signals $\rho \in [15,25]$ performing the worst comparatively. The accuracies of the $20-40M_\odot$ total mass slices are comparatively the lowest. (red dashed line $\equiv$ best network)}
\label{fig:bin_acc_slice}
\end{figure*}

The plots corresponding to the detection problem are in Figure~\ref{fig:bin_acc_slice}. All observations are consistent with NN output plots and slice ROC plots. The eccentricity slice $0-0.1$ has the lowest testing accuracy, since it forms the border between the two detection classes. All other slices have testing accuracies $>0.92$ after $9$ epochs. The accuracy across SNRs is almost uniform, with testing accuracies of all slices converging to $> 0.9$, with the faintest signals $\rho \in [15,25]$ performing the worst comparatively. The accuracies of the $20-40M_\odot$ total mass slices are comparatively the lowest.

\begin{figure*}[ht!]
\centering
\includegraphics{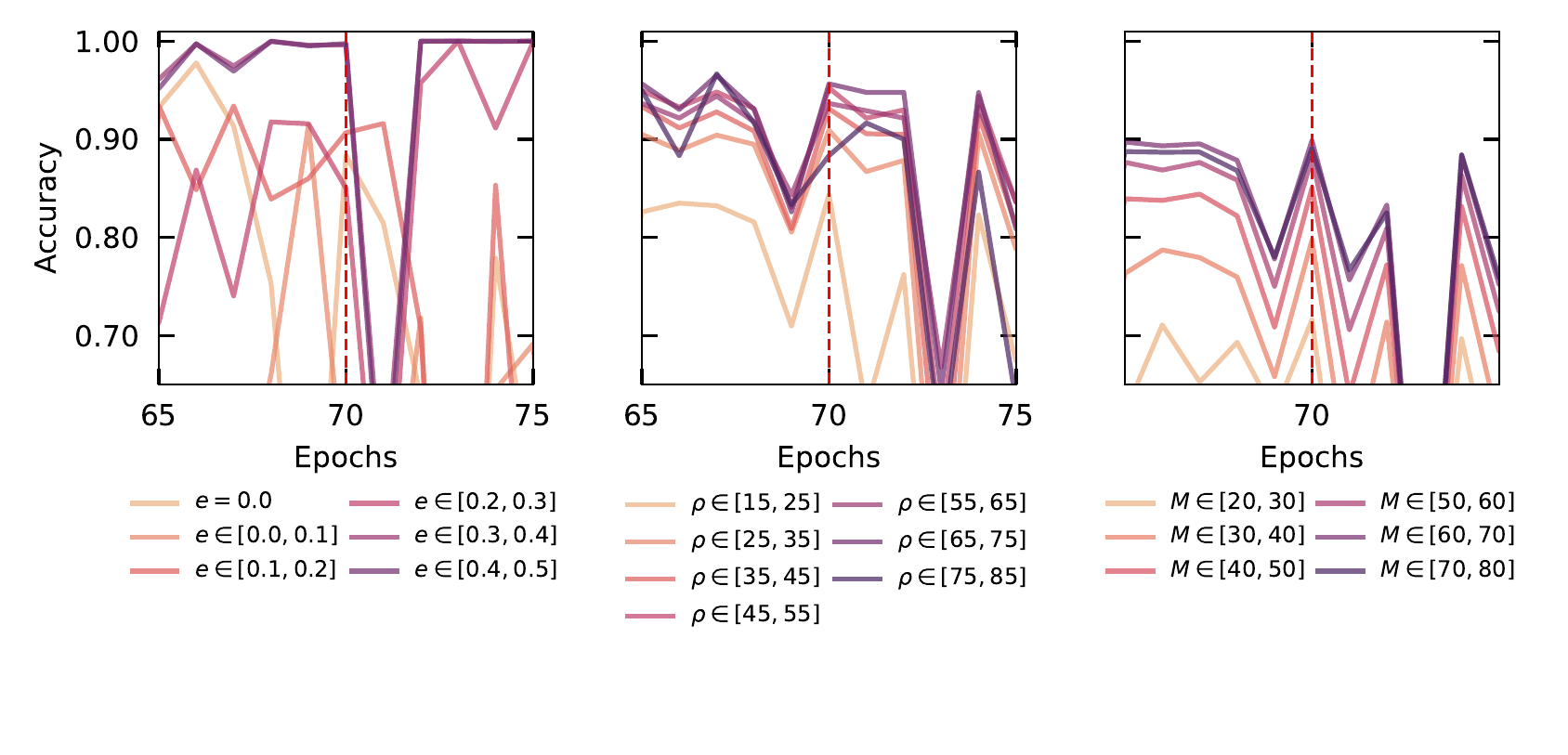}
\caption{Accuracies for the classification problem grouped by eccentricity (left), SNR (middle) and mass (right) ranges. The eccentricity slices $0-0.1$ and $0.2-0.3$ have the lowest testing accuracies, since they form the borders between the classes. We also continue to observe oscillations across the $100$ epoch, which does not seem to die out even at the onset of overfitting Figure~\ref{fig:tri_loss_acc}. But around the $80-90$ epoch range, these oscillations occur at $>0.9$ accuracy. The accuracies across SNR slices are almost uniform, with testing accuracies of all slices converging to $>0.94$. The loudest signals $\rho \in [65,85]$ are again the worst performing slice. The accuracy of the $20-30M_\odot$ total mass slice is again comparatively the lowest, but even this slice has a testing accuracy of $\sim 0.92$. The other slices have testing accuracies $>0.95$. (red dashed line $\equiv$ best network)}
\label{fig:tri_acc_slice}
\end{figure*}

The ones corresponding to the classification problem can be found in Figures~\ref{fig:tri_acc_slice}. The eccentricity slices $0-0.1$ and $0.2-0.3$ have the lowest testing accuracies, since they form the borders between the classes. We also continue to observe oscillations starting at $40$ epochs, which do not die out at higher epochs (Figure~\ref{fig:tri_loss_acc}). The accuracies across SNR slices are almost uniform with testing accuracies of all slices converging to $>0.9$, except the $\rho \in [15,25]$ slice. The faintest signals $\rho \in [15,25]$ are again the worst performing slice. The accuracy of the $20-40M_\odot$ total mass slice is again comparatively the lowest, going down to $.75$. The other slices have testing accuracies $>0.85$.

We further plot the ROCs of the best performing network, for subspaces (also referred to as slices) of {total mass} and {injection SNR}, to check for erroneous behaviors at different thresholds.

For the detection problem, in Figure~\ref{fig:bin_ROC}, we observe the $20-40M_\odot$ slices have ROCs with lowest TPRs. This lower performance can be attributed to the signal patterns being very long, hence making the pattern in the Q-scan either spread out beyond the range of our Q-scan. It can also be attributed to comparatively lower number of samples in the training set since the total mass distribution is a triangular distribution in $[20,80]$. Also, the faintest signals in the SNR range $\in [15,25]$ has the lowest TPRs, which can be seen as an indication of the fact that fainter patterns are more difficult for
the network to analyse and make predictions. $\rho \in [75,85]$ slice has the second lowest TPR, can be attributed to comparatively lower number of signals in that SNR range in a $\rho^{-4}$ distribution.

\begin{figure*}[ht!]
\centering
\includegraphics{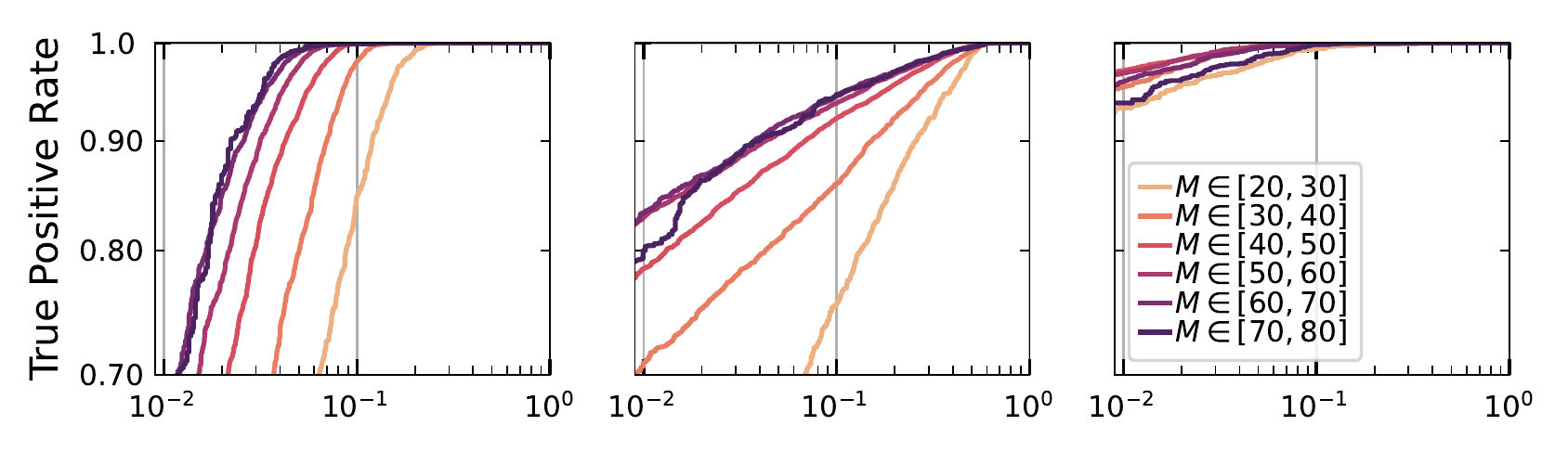}
\includegraphics{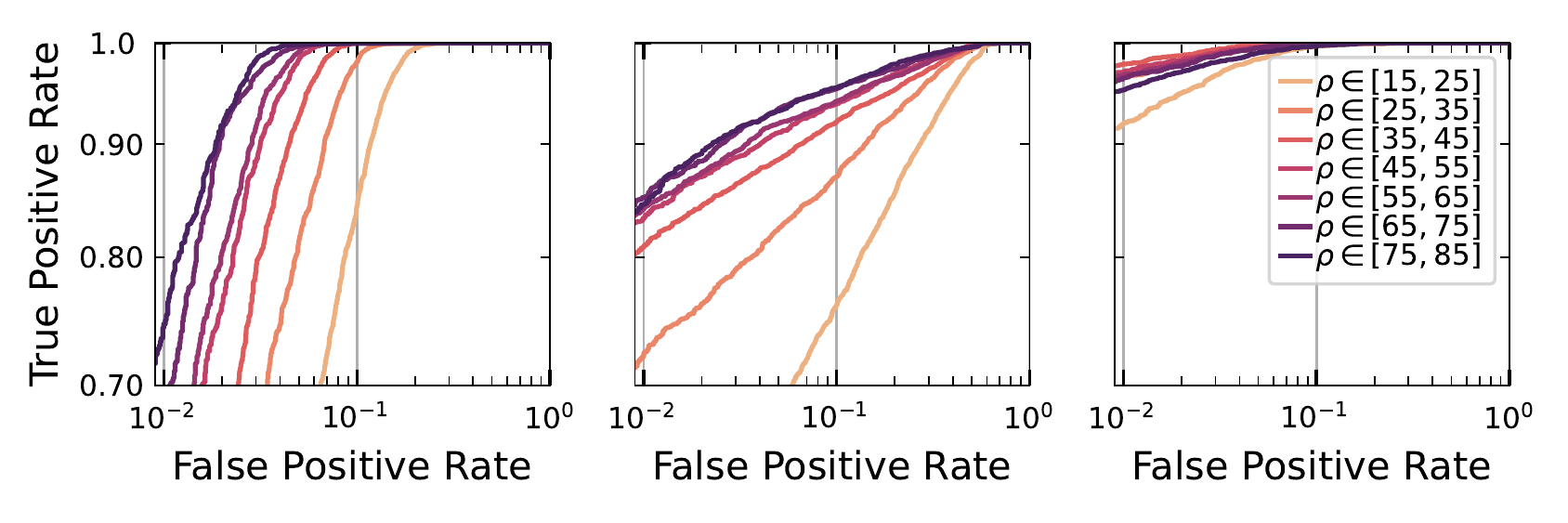}
\caption{ROC curves for the classification problem grouped by mass and SNR ranges. (Top) For FPRs as low as $0.1$, the TPRs are typically $>0.75$. The $20-40M_\odot$ slice has the lowest TPRs across all thresholds and across choices for positive class. (Bottom) For FPRs as low as $0.1$, the TPRs are $>0.75$. The faintest signals slice $\rho \in [15,25]$ is the worst performing slice.}
\label{fig:tri_ROC_slice}
\end{figure*}

For the classification problem, in Figure~\ref{fig:tri_ROC_slice}, for FPRs as low as $0.1$, the TPRs are as high as $>0.75$. The $20-40M_\odot$ slices have lowest TPRs across all thresholds and across choices for positive class. The faintest signals $\rho \in [15,35]$ again is the worst performing slice.

\section{Discussion and Future Work}\label{discussion}
In this work, as a proof of concept, we trained a separable convolutional neural network to distinguish eccentric GW signals from non-eccentric ones (detection problem), as well as to classify signals as non-eccentric, moderately eccentric, or highly/significantly eccentric (classification problem). Summary statistics for both problems  are described in Tables~\ref{tab:summary_det} and \ref{tab:summary_class} respectively. The cross-model accuracy tests indicate that waveform systematics do not impact network performance for the SNRs and detector networks considered in the work.

\begin{table}
\centering
\caption{Summary statistics for the best detection problem neural network, grouped by eccentricity, total mass, and SNR. The best performing network was defined as the one displaying lowest testing loss, which for the classification problem was the \texttt{real} trained network trained for $9$ epochs.}
\label{tab:summary_det}
\begin{tabular}{cc|cc|cc}
\toprule
    $e$ & Accuracy &   $M$ & Accuracy & $\rho$ & Accuracy \\
\midrule
      0 &    0.972 & 20-30 &    0.829 &  15-25 &    0.912 \\
  0-0.1 &    0.875 & 30-40 &    0.886 &  25-35 &    0.948 \\
0.1-0.2 &    0.918 & 40-50 &    0.915 &  35-45 &    0.964 \\
0.2-0.3 &    0.993 & 50-60 &    0.933 &  45-55 &    0.964 \\
0.3-0.4 &    0.999 & 60-70 &    0.945 &  55-65 &    0.979 \\
0.4-0.5 &    0.999 & 70-80 &    0.947 &  65-75 &    0.985 \\
        &          &       &          &  75-85 &    0.952 \\
\bottomrule
\end{tabular}
\end{table}

\begin{table}
\centering
\caption{Summary statistics for the best classification problem neural network,  grouped by eccentricity, total mass, and SNR. The best performing network was defined as the one displaying lowest testing loss, which for the classification problem was the \texttt{real} trained network trained for $70$ epochs.}
\label{tab:summary_class}
\begin{tabular}{cc|cc|cc}
\toprule
    $e$ & Accuracy &   $M$ & Accuracy & $\rho$ & Accuracy \\
\midrule
      0 &    0.883 & 20-30 &    0.716 &  15-25 &    0.844 \\
  0-0.1 &    0.531 & 30-40 &    0.797 &  25-35 &    0.910 \\
0.1-0.2 &    0.906 & 40-50 &    0.850 &  35-45 &    0.931 \\
0.2-0.3 &    0.850 & 50-60 &    0.881 &  45-55 &    0.952 \\
0.3-0.4 &    0.996 & 60-70 &    0.898 &  55-65 &    0.936 \\
0.4-0.5 &    0.998 & 70-80 &    0.890 &  65-75 &    0.956 \\
        &          &       &          &  75-85 &    0.883 \\
\bottomrule
\end{tabular}
\end{table}

We notice that samples on a real SNR data-set lead to better results, 
and hence we restrict to only the \texttt{real} trained networks in the discussion. The better performance was an indicator of the network requiring more samples in the low-SNR range to learn patterns in that subspace. This is because the neural network performs better when the distributions of the training set match the distributions of the test sets.

For the detection problem, the convergence to peak accuracy is fairly smooth, with a peak at $91.38\%$. Which in conjunction with all results listed in Section-\ref{results} indicates that the network architecture works well for the detection problem.

For the classification problem, however, the peak testing accuracy is comparatively lower ($85\%$); and the loss and accuracy curves are rapid oscillations that appear to never die out with epochs. Both these, together with the corresponding results in Section-\ref{results}, indicate that the current network architecture turns out to be insufficient for the classification problem. After $40$ epochs, the optimizer tries to vary the jump sizes on the loss surface in order to find a better global minimum, which doesn't seem to exist, thereby causing the large oscillations. Improving performance on the classification problem would require a network of higher complexity.

% , for which which reducing oscillations would be simple enough to achieve using an explicit learning rate decay.

% We observed oscillations in the initial epochs. The reasons for the same could be one or a combination of the following:
% \begin{enumerate}
%     \item The loss surface for the binary-cross-entropy function happens to be extremely uneven, hence initial gradient descent iterations at initial epochs cause large changes in the loss/accuracy.
%     \item It could also be the case that the initial starting point (initial learnable parameter values) lies in an uneven region of the loss function, and the gradient descent iterations move the NN state to a more uniform region of the loss function as iterations go by.
%     \item It could also be that the initial learning rate is high for the binary problem, which leads to big jumps on the loss surface; since Adam's optimizer brings the learning rate down to a smaller value as iterations go by, the size of the jumps on the loss surface reduces and the testing accuracy line becomes smoother.
% \end{enumerate}

% However, this is generally expected while training a neural network, and is not a cause for concern as long as we stop our training well after these oscillations take place, which we have done. It also has no bearing on the trained network one finally uses.

\begin{figure}
    \centering
    \includegraphics{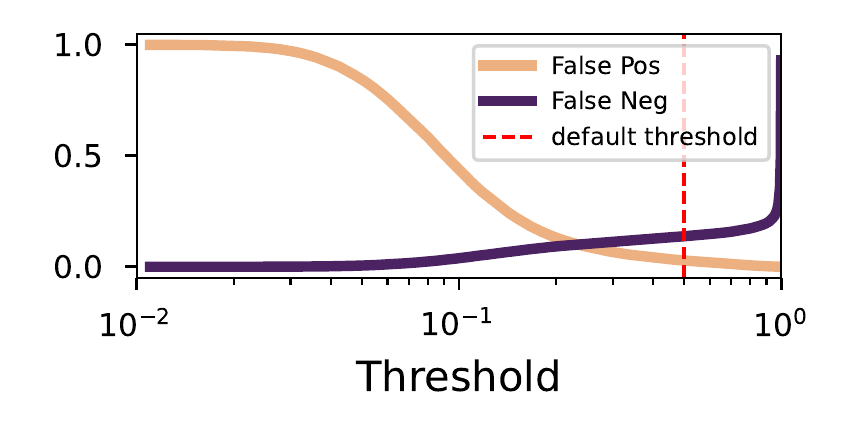}
    \caption{The False Positive Rate and False Negative Rates plotted as a function of the threshold for the detection network (where red dashed line is the default threshold). One can decrease threshold to decrease false negative rate at the cost of increasing the false positive rate. False negatives are extremely undesirable since the correspond to eccentric signals that were falsely labeled as non-eccentric and will be missed in downstream analysis. False positives correspond to non-eccentric signals that are falsely labeled as eccentric that will be passed on for downstream analysis and increase computational cost there.}
    \label{fig:fpr_fnr_thresh}
\end{figure}

Confusion Matrices are extremely close to identity matrices of the appropriate dimension for the detection problem. This indicates proximity to the behavior of an ideal detector. Most results on the binary classifier corresponded to the default threshold value of $0.5$. But if one requires the detector to have an high true positive rate at the cost of a non-negligible false-positive-rate, (which would correspond to detecting nearly all true eccentric signals even if some non-eccentric signals are incorrectly detected), the threshold value could be decreased accordingly (Figure-\ref{fig:fpr_fnr_thresh}). The degree to which it could be decreased can be estimated from the ROC curves in Figures~\ref{fig:bin_ROC}, \ref{fig:tri ROC} and \ref{fig:tri_ROC_slice}, based on the tolerance that one sets for the False Positive Rate.

For the detection problem slice accuracy tests, the eccentricity slice $e \in (0, 0.1)$ has a comparatively poorer behavior. This can be understood because essentially the boundary between the eccentric and non-eccentric classes, and hence these signals are the most difficult to classify correctly. The total-mass slices in $(20M_\odot ,30 M_\odot)$ have the poorest performance. This is because the total length of a signal scales inversely with the total mass, and hence less massive systems produce longer signals. This implies that the information is spread over a longer duration, which would require longer Q-scan images to be trained on the network. Hence these are the most difficult to classify correctly. The performance of all injection SNR slices was almost the same, with the faintest ($[15-25]$ SNR) signals have comparatively the lowest accuracy. These observations on comparative performance were backed up by the slice ROC test plots \ref{fig:bin_ROC}.

Additionally for the detection problem, one can decrease neural network threshold to decrease false negative rate at the cost of increasing the false positive rate. (Figure~\ref{fig:fpr_fnr_thresh}) False negatives correspond to eccentric signals that were falsely predicted as non-eccentric and hence will be missed in downstream analysis. False positives correspond to non-eccentric signals that are falsely predicted as eccentric that will be passed on for downstream analysis. The former is extremely undesirable since that would imply some eccentric signals won't be analysed as eccentric thereby introducing systematic biases in estimating properties of that signal's source. The latter would imply some (nearly-)quasi-circular signals will be analysed as eccentric result of increasing the computational cost in downstream analysis.

For the classification problem slice accuracy tests, the eccentricity slices $(0, 0.1)$ and $(0.2, 0.3)$ have comparatively poorer behavior. This is because these slices form the boundaries between classes, and hence these signals are the most difficult to classify correctly. The total-mass slices in $(20M_\odot ,30 M_\odot)$ again have the poorest performance. The lowest SNR slices again had comparatively the lowest accuracy. These observations on comparative performance were backed up by the slice ROC test plots in Figure~\ref{fig:tri_ROC_slice}.

If advanced training procedures, like curriculum learning, are to be implemented on similar network architectures for similar tasks, the slice accuracy and ROC curves can be referred to for information on which parameter-subspaces contain ``difficult'' samples for the network to classify. In particular, it was observed that the samples near the eccentricity class boundaries had the poorest performance, and hence could be labeled ``difficult'' for the network to learn. What could also help is modifying the training set parameter distributions to have more samples from ``difficult to learn'' parameter subspaces.

Testing our network on real catalogue events is beyond the scope of this paper, but is certainly an avenue for future work. Specifically, since we trained on synthetic Gaussian noise using O4 representative \texttt{PSD}s, our current network may not be optimal for real events. Additionally, the real events would also require the training data to include spin, which hasn’t been included in this work.

We end by mentioning avenues for future work:
\begin{enumerate}
\item Improving performance on the classification problem by implementing a network of higher complexity.
\item Extending this work to build a regressive network to extract the numerical value of the eccentricity to a certain accuracy. 
\item We also plan to extend the parameter ranges of the GW signals, in particular the binary masses and component binary spins. But since the length of the signal roughly scales with the total mass, the NN architecture will need to be modified to accommodate the longer and shorter Q-scans of less and more massive binary systems respectively.
\item At the moment, our training samples were embedded in Gaussian noise generated using representative power-spectral-densities. A possible avenue is to use real detector noise to train the samples on noise that the detectors actually record.
\item As the \texttt{EccentricTD} waveform template extends to include spins of the binary objects, there will be some changes in the Q-transforms. And the NNs will have to be retrained on the newly generated samples. The currently trained NN could be used at the base network for that problem.
\item Since it is known that there might be some degeneracy between eccentric and precessing signals (see eg.~\cite{OShea:2021ugg,CalderonBustillo:2020xms}, training a modified network, using this as the base architecture, to distinguish between the two could be a topic of future work.
\item As more GW-detectors become operational, the network will have to be retrained using augmented training data, in particular, the number of channels of the input images will increase and each new channel will correspond to a new detector. As for the network architecture, only the first layer will have to be modified to accept images with more than $3$ channels as detectors \cite{gebhard2019convolutional}.
\end{enumerate}

\section*{ACKNOWLEDGEMENTS}\label{acknowledgements}
We thank Shreejit Jadhav for carefully reviewing this work, and Viola Sordini for useful comments. We also thank Anupreeta More and Isobel M Romero-Shaw for their comments. This work is supported by the Department of Atomic Energy, Government of India, under Project No. RTI4001. AV is also supported by a Fulbright Program grant under the Fulbright-Nehru Doctoral Research Fellowship, sponsored by the Bureau of Educational and Cultural Affairs of the United States Department of State and administered by the Institute of International Education and the United States-India Educational Foundation. PK is also supported by Ashok and Gita Vaish Early Career Faculty Fellowship at ICTS. 
AR is also partially supported by the Department of Mathematics, University of Massachusetts, Dartmouth.
We thank the International Centre for Theoretical Sciences, Tata Institute of Fundamental Research, Bangalore; for access to their computing cluster, which was used for all computing in this work.
This work was presented at the (bi-)weekly meetings of LIGO CBC-R-D subggroup and the LIGO-India Scientific Collaboration. We thank the attendees of those meeting for their inputs and suggestions. 

\section*{References}
\bibliography{references}

\end{document}